\documentclass[a4paper,12pt,pdftex]{article}
\usepackage[utf8]{inputenc}

\usepackage{amsmath}
\usepackage{amsthm}
\usepackage{amssymb}
\usepackage{hyperref}
\usepackage{caption}
\usepackage{float}
\usepackage{color}
\usepackage{graphicx}
\floatstyle{plaintop}
\restylefloat{figure}
\usepackage{amssymb}
\usepackage{subcaption}
\usepackage[margin=0.5in]{geometry}
\title{Resource requirements and speed versus geometry of unconditionally secure physical key exchanges}
\author{Elias Gonzalez $^{1,}$*, Robert S. Balog $^{1}$ and Laszlo B. Kish $^{1}$ }
\date{$^{1}$ Texas A\&M University Department of Electrical and Computer Engineering, 3128 TAMU College Station, TX 77843, U.S.A. \\ * corresponding author eliasg23@tamu.edu}

\begin{document}

\maketitle

\begin{abstract}
The imperative need for unconditional secure key exchange is expounded by the increasing connectivity of networks and by the increasing number and level of sophistication of cyberattacks. Two concepts that are information theoretically secure are quantum key distribution (QKD) and Kirchoff-law-Johnson-noise (KLJN). However, these concepts require a dedicated connection between hosts in peer-to-peer (P2P) networks which can be impractical and or cost prohibitive. A practical and cost effective method is to have each host share their respective cable(s) with other hosts such that two remote hosts can realize a secure key exchange without the need of an additional cable or key exchanger. In this article we analyze the cost complexities of cable, key exchangers, and time required in the star network. We mentioned the reliability of the star network and compare it with other network geometries. We also conceived a protocol and equation for the number of secure bit exchange periods needed in a star network. We then outline other network geometries and trade-off possibilities that seem interesting to explore.
\end{abstract}

\section{Introduction}

\subsection{Motivation for a secure network}

In the advent of intelligent vehicle information networks \cite{evs}, the smart power grid \cite{powergrid}, and the Internet of Things (\textit{IoT}) \cite{iot}, current infrastructure is becoming increasingly dependent on cyber networks. This dependency makes current infrastructure a larger more attractive target for cyberattacks, such that the National Security Agency (NSA) director stated the U.S. power grid could be shut down with a cyberattack~\cite{nsa_china}. 

\begin{figure}[h]
    \caption{An illustration of Alice and Bob in a secure key exchange while Eve is seeking to tap the communication channel and extract the key.}
    \label{fig:alice_bob}
  \centering
    \includegraphics[width=0.75 \textwidth]{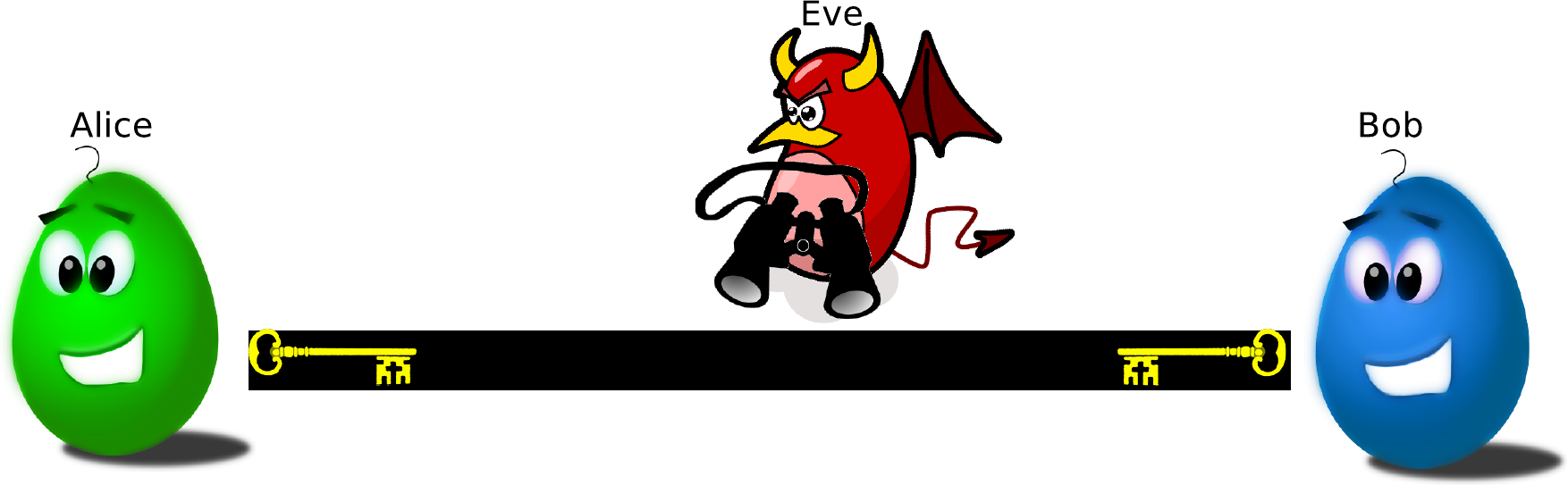}
\end{figure}

Secure communication channels are needed to prevent eavesdropping or intervention. Increasingly though, communications is directed away from expensive, dedicated networks in favor of the open internet. In order to ensure secure communications, security keys are needed to set up a secure communication.  The keys are generated, and shared via a publicly accessible channel by secure key distribution protocols. Consider a secure key exchange between Alice and Bob, Alice and Bob must consider that an eavesdropper (Eve) is trying to extract the key as illustrated in Figure~\ref{fig:alice_bob}. Secure key exchanges can be categorized as either software-based or hardware-based.

Software-based key exchanges are based on mathematical algorithms with the assumption that Eve does not have enough computing resources to crack the key. In essence, software-based key exchanges offer no security from an information theoretical point of view. The security is only (computationally-) conditional and is not \textit{future-proof}, meaning that with enough computing resources the key can be extracted. The advantages of software-based key exchanges are the low cost, hardware communicator is not required, and the keys can be exchanged over the Internet, thus eliminating the need of extra infrastructure. The other option is hardware-based key exchange, these offer an advantage of unconditional security. 

\subsection{Hardware-based secure key exchanges}

The Quantum Key Distribution (QKD) \cite{bb84} and the Kirchhoff-Law-Johnson-Noise (KLJN) \cite{kljn1, lk1, lk2, lk3, kljn_its, kljn_noise_properties, kljn_noise_properties2, barry1, barry2, c186, c174, c171, c169, c147, c141, c133, c128, c118, c111, c113, c+6} secure key exchange are two examples of hardware-based secure key exchange concepts that are information theoretically secure \cite{infotheosec}. Thus even with infinite computing resources the key will not be extracted by Eve, because the security offered by these schemes are based on fundamental laws of physics, to crack the key exchange would require Eve to break the underpinning laws of physics. The main disadvantage of hardware-based key exchanges is the higher cost, as they require a physical communicator at each host, and a dedicated connection between communicators. Such communication schemes can be considered peer-to-peer (P2P) \cite{def_p2p}.

The QKD key exchange utilizes the quantum no-cloning theorem of quantum mechanics \cite{bb84} to distribute key bits. In theory it is information theoretically secure, however the physical implementation of QKD has been debated and the method has been hacked \cite{yuen, crack2, crack3, crack4}.

\begin{figure}[h]
    \caption{An illustration of a KLJN system. Alice and Bob each have a communicator which have noise generators, a low resistor $R_\mathrm{L}$, and a high resistor $R_\mathrm{H}$. The noise voltages are enhanced by Johnson noise $U_\mathrm{A,L}$ or $U_\mathrm{A,H}$ for Alice; and $U_\mathrm{B,L}$ or $U_\mathrm{B,H}$ for Bob, which is measured between the wire and the ground. Once the communicators select a resistor they measure the mean-squared voltage amplitude $<U^{2}_\mathrm{ch}(t)>$ and or the current amplitude $<I^{2}_\mathrm{ch}(t)>$.}
    \label{fig:kljn_sys}
  \centering
    \includegraphics[width=0.75 \textwidth]{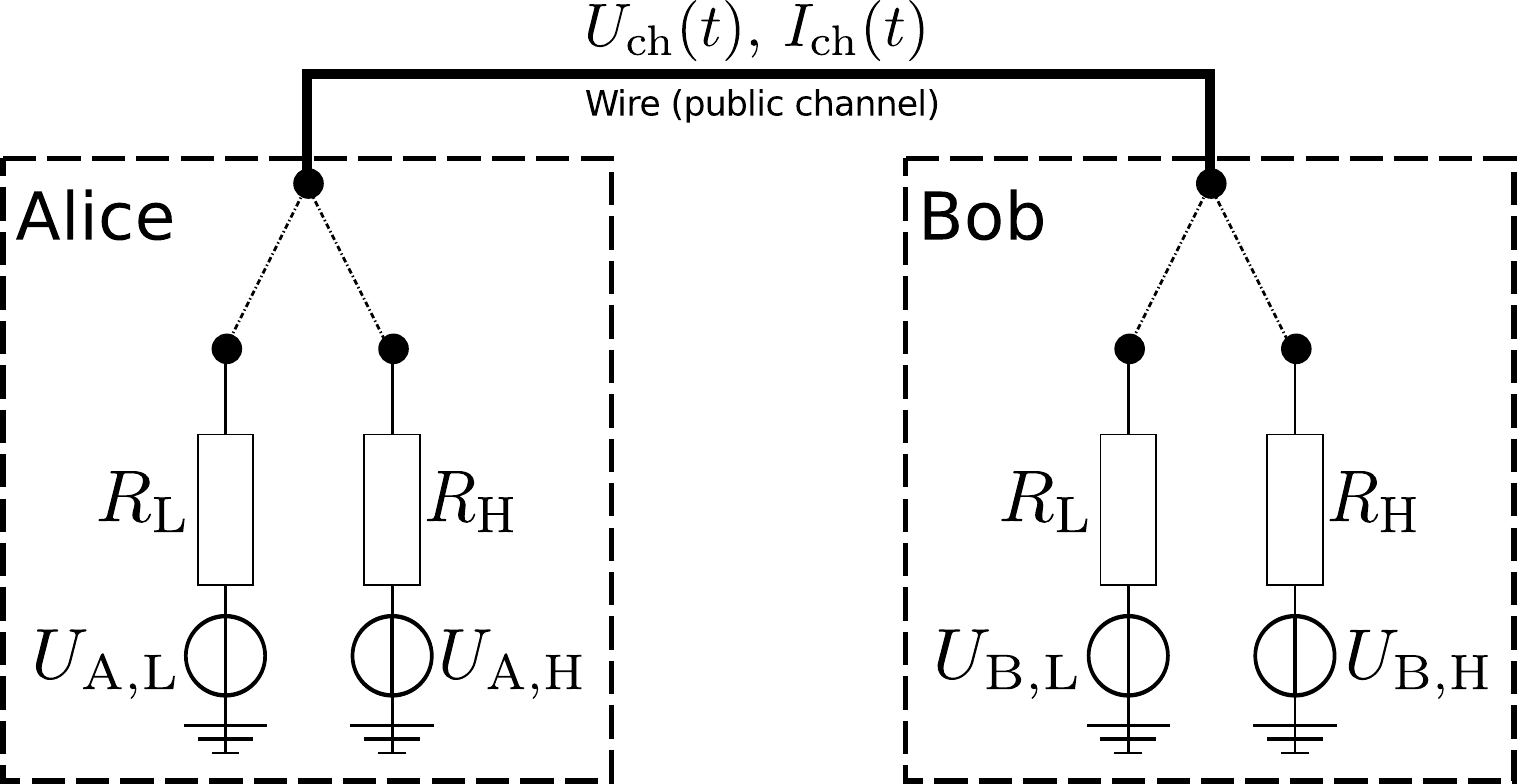}
\end{figure}

The KLJN key exchange utilizes the laws and properties of classical mechanics \cite{kljn1} to generate and distribute key bits. In the KLJN key exchange depicted in Figure~\ref{fig:kljn_sys}, Alice and Bob have two identical resistor pairs, $R_\mathrm{L}$ and $R_\mathrm{H}$ (the values of the resistors are such that $R_\mathrm{L}$ < $R_\mathrm{H}$) which represents the low $R_\mathrm{L}$ and high $R_\mathrm{H}$ bits respectively. At the beginning of each bit exchange period, the communicators randomly generate a bit value and connect the corresponding resistor to the wire line. The effective value of the resulting thermal noise in the cable has three possible levels. When it is at the intermediate level, Alice and Bob will know that the other party has the opposite bit value than their own. Thus a secure bit exchange took place because Eve, while she also knows that Alice and Bob have opposite bit values, she does not know who has the $R_\mathrm{L}$ value and who has the $R_\mathrm{H}$ value \cite{kljn1, lk2}.

\subsection{Secure key exchange over P2P networks and the fully connected network}

Hardware-based key exchanges require P2P networks with a dedicated connection to each host. For very large networks this will be costly due to the infrastructure (cables) and key exchangers. The cost complexity of the growth for different networks can be denoted by $T_\mathrm{cable}(N)$ for number of cables, $T_\mathrm{ke}(N)$ for number of key exchangers, and $T_\mathrm{time}(N)$ for amount of time required or speed to complete a secure bit exchange, with $N$ representing the number of hosts in the network. 

\begin{figure}[h]
    \caption{An illustration of a fully connected network with $N-1$ communicators per host (denoted as $\mathrm{FCN}_{N-1}$) has complexities of $T_\mathrm{cable}(N) \in O(N^{2})$, $T_\mathrm{ke}(N) \in O(N^{2})$, and $T_\mathrm{time}(N) \in O(1)$. }
    \label{fig:bf1}
  \centering
    \includegraphics[width=0.5 \textwidth]{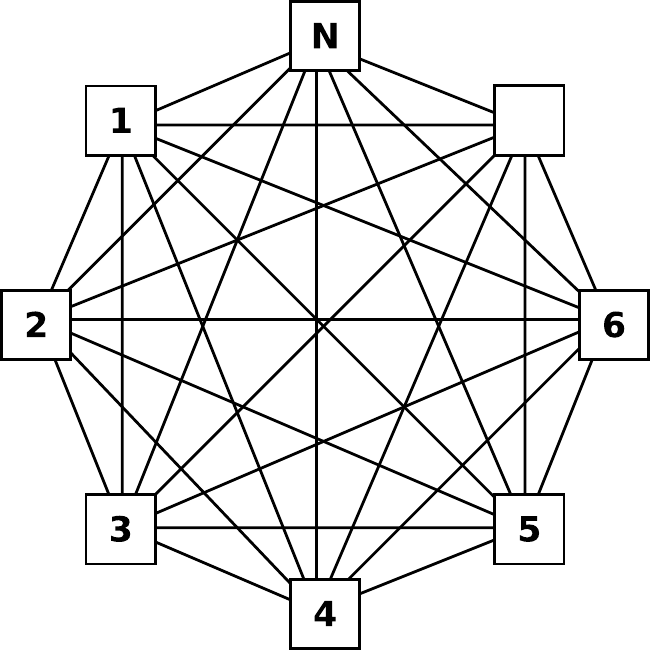}
\end{figure}

A simple method to construct P2P networks is a fully connected network also known as the complete graph in graph theory. The fully connected network is illustrated in Figure~\ref{fig:bf1}. The fully connected network does not require a protocol since every host in the network has a dedicated connection with every other host in the network, and can process a secure bit exchange with any other host at any time simultaneously. This network has $N-1$ key exchangers per host and scales with the order of $N^{2}$ for cables and key exchangers, which makes this network impractical for very large networks. The complexities are $T_\mathrm{cable}(N) \in O(N^{2})$, $T_\mathrm{ke}(N) \in O(N^{2})$, and $T_\mathrm{time}(N) \in O(1)$. We will denote the fully connected network with $N-1$ key exchangers per host as $\mathrm{FCN}_{N-1}$. The fully connected network has $N-1$ key exchangers for every host resulting in $(N-1) \cdot N$ total key exchangers for the entire network, $N-1$ direct connections for every host resulting in $(N-1) \cdot N/2$ total cables for the entire network. The advantage the fully connected network has is time, as every host in the network can simultaneously process a secure bit exchange with every other host in the network.

If the cost of having $(N-1) \cdot N$ key exchangers for the entire network is too costly, then a trade-off between the number of key exchangers and speed might be preferable. If there is only one key exchanger per host in the fully connected network then the complexities for the fully connected network  will be; $T_\mathrm{cable}(N) \in O(N^{2})$, $T_\mathrm{ke}(N) \in O(N)$, and $T_\mathrm{time}(N) \in O(N)$, and will require a protocol which we will denote as $\mathrm{FCN}_{1}$ to process a secure bit exchange with every host in the network.

The fully connected network is robust and reliable as it does not depend on a single cable or key exchanger. If there is cable destruction or a damaged key exchanger then only the hosts connected by that cable or key exchanger will be affected, and only that connection will be affected. The affected hosts will still be able to process a secure bit exchange with other hosts which do not depend on the damaged cable or key exchanger. 

To add additional hosts to the fully connected network will be trivial since it does not have a protocol. In the case of $\mathrm{FCN}_{1}$ the protocol will need to consider the added host.

\subsection{Linear chain network with two key exchangers per host}

\begin{figure}[h]
    \caption{An illustration of a linear chain network with $2$ key exchangers per host has complexities of $T_\mathrm{cable}(N) \in O(N)$, $T_\mathrm{ke}(N) \in O(N)$, and $T_\mathrm{time}(N) \in O(N^{2})$. }
    \label{fig:lch}
  \centering
    \includegraphics[width=0.75 \textwidth]{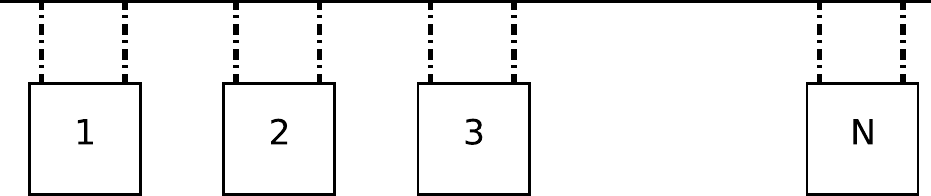}
\end{figure}

Linear chain networks also know as bus networks or daisy chain networks, contain a single line and two key exchanges per host as illustrated in Figure~\ref{fig:lch}, and were analyzed in \cite{me1} in the contexts of smart grids. The linear chain network with $2$ key exchangers per host has complexities of $T_\mathrm{cable}(N) \in O(N)$, $T_\mathrm{ke}(N) \in O(N)$, and $T_\mathrm{time}(N) \in O(N^{2})$. By having $2$ key exchanges per host the linear chain network can process $2$ simultaneous secure bit exchanges as long as one host is downstream, say host \textit{i}$-a$ for any positive integer $a$ and the other host is upstream, say host \textit{i}$+b$ for any positive integer $b$ of the \textit{i}th host. The first host and the last host are special cases which cannot have simultaneous secure key exchanges with other hosts \cite{me1}. 

The reliability of the linear chain network is dependent on the cable. If there is damage to the cable then the network will become two different networks divided at the location of the damaged cable, and the two networks cannot process a secure bit exchange with each other. The linear chain network is more robust if there is damage to a key exchanger, then only the host with the damaged key exchanger will be slowed down but will be able connect with all other hosts in the network since there are two key exchangers per host. 

If an additional host joins the network with $N$ hosts then the protocol will consider $N+1$ hosts instead of $N$, this will be a relatively simple fix as the the protocol can be preprogrammed in the hosts for any $N$.

\section{Results and Discussion}

\subsection{Star network}

\begin{figure}[h]
    \caption{An illustration of a star network system with one key exchanger per host has complexities of $T_\mathrm{cable}(N) \in O(N)$, $T_\mathrm{ke}(N) \in O(N)$, and $T_\mathrm{time}(N) \in O(N)$. }
    \label{fig:starn}
  \centering
    \includegraphics[width=0.5 \textwidth]{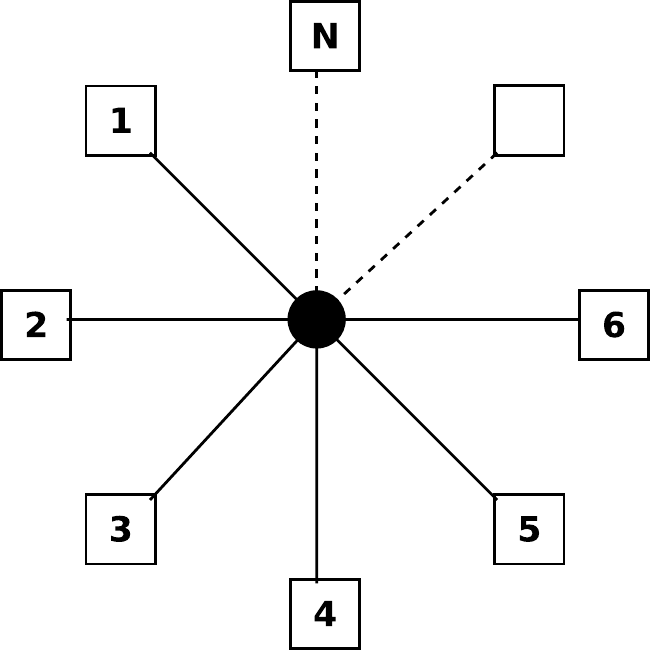}
\end{figure}

The star network is a hub and spoke topology with a center switch like an old telephone exchange switch system, and has branches connected to the center. We denote the star network protocol with one key exchanger per host as $\mathrm{STAR}$. The complexities of the star network are $T_\mathrm{cable}(N) \in O(N)$, $T_\mathrm{ke}(N) \in O(N)$, and $T_\mathrm{time}(N) \in O(N)$. Figure~\ref{fig:starn} is an example of a star network with $N$ branches.

The most efficient protocol in the star network is similar to the protocol in the linear chain network in regards to first connecting to the nearest neighbors, then connecting the second nearest neighbors, and so on. The star network allows for faster speed than the linear chain network with similar cable and hardware complexities.

\subsection{Graph theory and previous work on the star network}

In graph theory the hosts are considered vertices and the cables are considered edges~\cite{introgt}. The protocol of the star network is to connect every host in the network to process a secure bit exchange with every other host in the network in the least number of Secure Bit Exchange Period (SBEP) steps. In graph theory the star network protocol can be described as a special case of a edge-color problem~\cite{edgecolor} known as round-robin(RR) tournament or all-play-all tournament problem~\cite{rrp}. The number $k$ of edge colors needed in graph theory is the number of SBEPs needed in the star network protocol, although many geometric structures and edge-color problems have been studied in graph theory~\cite{tutte1, tutte2, best, best2, book1, book2, book3} and applied to various infrastructure networks~\cite{apply1, apply2, apply3, book4}, it has not been applied to P2P hardware-based secure key exchange networks other than~\cite{me1}. Many network applications assume overlapping signals in the same channel is possible, and do not have a dedicated channel in which every vertex connects with every other vertex. For QKD and KLJN network applications these networks require dedicated communication channels with no overlapping signals, and RR solutions to different geometric structures. The star network protocol presented in section 2.2 is specifically for QKD and KLJN networks, and is significant since it combines residual SBEP steps whenever possible, thus lowering the total number of SBEPs needed, after a thorough literature review a similar RR solution was not found and the most similar solution found is in~\cite{best}.

\subsection{Protocol and analysis of the star network}

For a network with $N$ hosts the star key exchange network protocol begins with every odd numbered host say \textit{i}th host with \textit{i} being odd and processes a secure bit exchange with their upstream nearest neighbor, that is host \textit{i}+1, this will take one Secure Bit Exchange Period (SBEP) and the secure key exchange between different hosts will occur simultaneously. For example, host $1$ will process a secure bit exchange with host $2$, while host $3$ will process a secure bit exchange with host $4$, while host $N-1$ will process a secure bit exchange with host $N$ if $N$ is even, or host $N-2$ will process a secure bit exchange with host $N-1$ if $N$ is odd. If $N$ is odd, then the last host, that is host $N$, will not process a secure bit exchange in the first SBEP step. The next step in the protocol is for every even numbered host say \textit{i}th host with \textit{i} being even will process a secure bit exchange with their nearest upstream neighbor, say host \textit{i}+1, simultaneously. For example, host $2$ will process a secure bit exchange with host $3$, while host $4$ will process a secure bit exchange with host $5$, while host $N-1$ will process a secure bit exchange with host $N$ if $N$ is even, or host $N$ will process a secure bit exchange with host $1$ if $N$ is odd, note that the protocol will wrap around from the last host $N$ to the first host $1$. The circular nature of the star network is a reason why it is faster than the linear chain network with similar cable and hardware complexities. The star network protocol $\mathrm{STAR}$ then continues with every odd host to process a secure bit exchange with their upstream second nearest neighbor, that is every \textit{i}th host with \textit{i} being odd with host \textit{i}+2, then the even numbered hosts will process a secure bit exchange with their second nearest neighbor, say every \textit{i}th host with \textit{i} being even with host \textit{i}+2. The protocol continues by having every host process a secure bit exchange with their third nearest neighbors, then fourth nearest neighbors, and continues until every host in the network has processed a secure bit exchange with every other host.

\begin{figure}[t]
        \centering
        \begin{subfigure}[b]{0.23\textwidth}
		\caption{1\textit{st} SBEP step}
                \includegraphics[width=\textwidth]{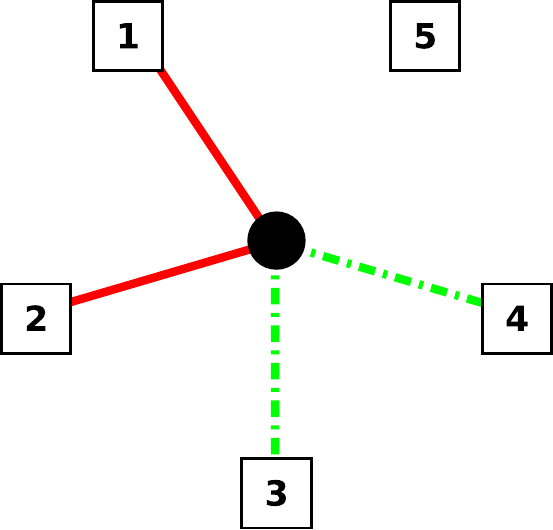}
                \label{fig:step1}
        \end{subfigure}%
        \hfill 
        \begin{subfigure}[b]{0.23\textwidth}
		\caption{2\textit{nd} SBEP step}
                \includegraphics[width=\textwidth]{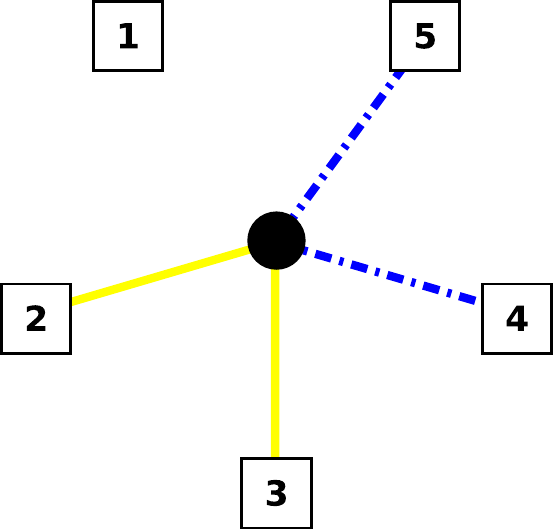}
                \label{fig:step2}
        \end{subfigure}
        \hfill 
        \begin{subfigure}[b]{0.23\textwidth}
		\caption{3\textit{rd} SBEP step}
                \includegraphics[width=\textwidth]{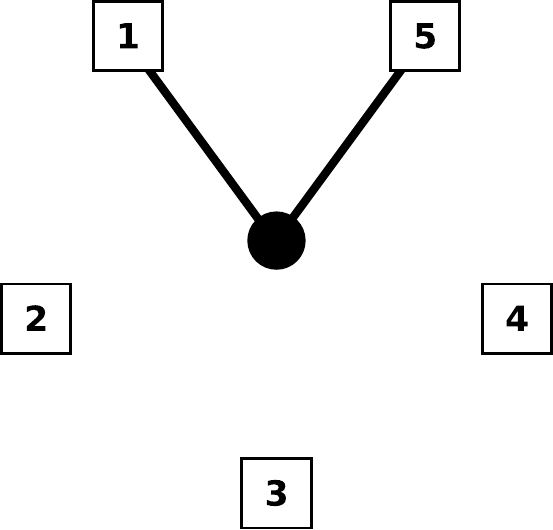}
                \label{fig:step3}        
        \end{subfigure}
        \\
        \begin{subfigure}[b]{0.23\textwidth}
		\caption{4\textit{th} SBEP step}
                \includegraphics[width=\textwidth]{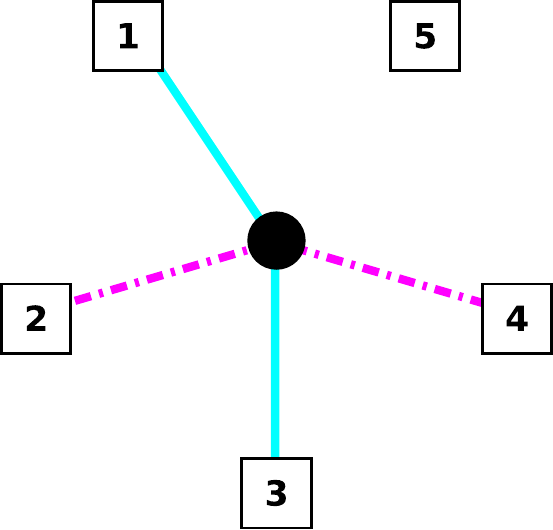}
                \label{fig:step4}
        \end{subfigure}%
        \hfill 
        \begin{subfigure}[b]{0.23\textwidth}
		\caption{5\textit{th} SBEP step}
                \includegraphics[width=\textwidth]{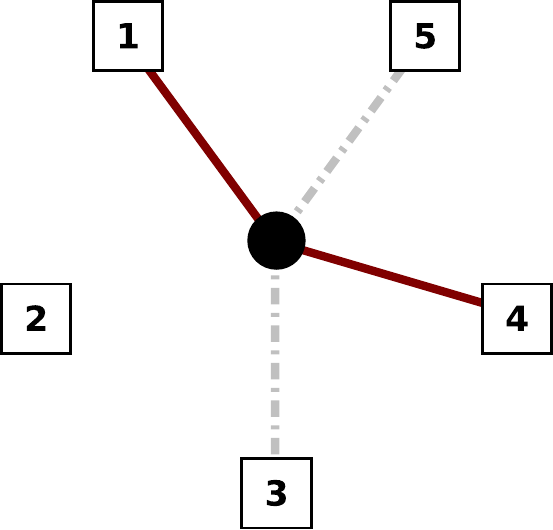}
                \label{fig:step5}
        \end{subfigure}
        \hfill 
        \begin{subfigure}[b]{0.23\textwidth}
		\caption{6\textit{th} SBEP step}
                \includegraphics[width=\textwidth]{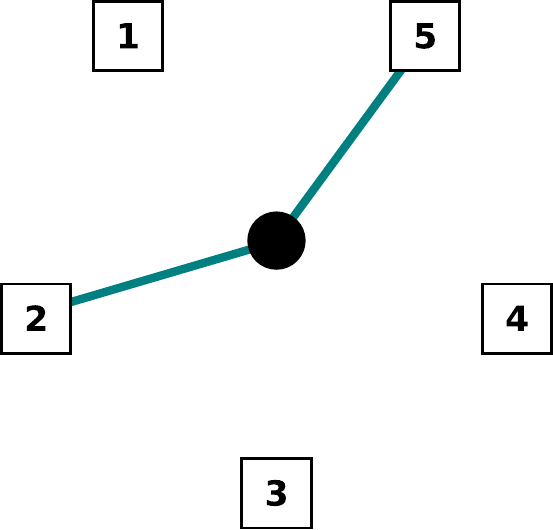}
                \label{fig:step6}        
        \end{subfigure}
        \caption{An illustration of the example of the star network protocol $\mathrm{STAR}$ for a network with $5$ hosts. It takes six SBEP steps for every host in the network to process a secure bit exchange with every other host.}
        \label{fig:steps}
\end{figure}

As an example Figure~\ref{fig:steps} illustrates every step of the protocol STAR for a network with 5 hosts. The first SBEP step in the protocol is illustrated in sub-figure~\ref{fig:step1}, note how every odd numbered host \textit{i} has a secure bit exchange with their next upstream nearest neighbor host \textit{i}+1. The second SBEP step in the protocol is illustrated in sub-figure~\ref{fig:step2}, note how every even numbered host \textit{i} has a secure bit exchange with their next upstream nearest neighbor host \textit{i}+1. The third SBEP step in the protocol is illustrated in sub-figure~\ref{fig:step3}. Since the number of hosts in the network is odd it will take additional SBEP steps to process a secure bit exchange with these remaining hosts, these are residual SBEP steps. Note how the last host wraps around to the first host. The fourth SBEP step in the protocol is illustrated in sub-figure~\ref{fig:step4}. In this SBEP step every odd numbered host \textit{i} has a secure bit exchange with their second upstream nearest neighbor host \textit{i}+2. The fifth SBEP step in the protocol is illustrated in sub-figure~\ref{fig:step5}, this step is similar to step 4 except that now the even numbered hosts process a secure bit exchange with their second upstream nearest neighbors. The sixth and last SBEP step in the protocol is illustrated in sub-figure~\ref{fig:step6}. Since $N$ is odd the protocol requires additional residual SBEP steps to process a secure bit exchange with the remaining hosts. Note that for this example of the STAR protocol with $N=5$ hosts requires six SBEP steps for every host in the network to process a secure bit exchange with every host. Table~\ref{table:steps} demonstrates what every host is doing at every step in the protocol of this example as illustrated in Figure~\ref{fig:steps}. Table~\ref{table:legend} is the legend for table~\ref{table:steps}. The arrow symbol ``$\rightarrow$'' is used as $x \rightarrow y$ meaning host $x$ is processing a secure bit exchange with host $y$. The star symbol ``$\bigstar$'' means the host of this row is being utilized. The circle symbol ``$\bigcirc$'' means the host of this row is not active.

\begin{table}[H]
  \begin{center}
    \begin{tabular}{| c | c | c | c | c | c | c |}
    \hline
    Host & (a) 1\textit{st} SBEP & (b) 2\textit{nd} SBEP & (c) 3\textit{rd} SBEP & (d) 4\textit{th} SBEP & (e) 5\textit{th} SBEP & (f) 6\textit{th} SBEP \\
    \hline
     1 & 1 $\rightarrow$ 2 & $\bigcirc$ & $\bigstar$ & 1 $\rightarrow$ 3 & $\bigstar$ & $\bigcirc$ \\
     \hline
     2 & $\bigstar$ & 2 $\rightarrow$ 3 & $\bigcirc$ & 2 $\rightarrow$ 4 & $\bigcirc$ & $\bigstar$ \\
     \hline
     3 & 3 $\rightarrow$ 4 & $\bigstar$ & $\bigcirc$ & $\bigstar$ & 3 $\rightarrow 5$ & $\bigcirc$ \\
     \hline
     4 & $\bigstar$ & 4 $\rightarrow$ 5 & $\bigcirc$ & $\bigstar$ & 4 $\rightarrow$ 1 & $\bigcirc$ \\
     \hline
     5 & $\bigcirc$ & $\bigstar$ & 5 $\rightarrow$ 1 & $\bigcirc$ & $\bigstar$ & 5 $\rightarrow$ 2 \\
    \hline
    \end{tabular}
  \end{center}
  \caption{This table demonstrates what every host is doing at every SBEP step in the protocol $\mathrm{STAR}$ as described in the example and illustrated in Figure~\ref{fig:steps}.}
  \label{table:steps}
\end{table}

\begin{table}[H]
  \begin{center}
    \begin{tabular}{| c | c |}
    \hline
    Symbol & Meaning of symbols in table~\ref{table:steps}\\
    \hline
    $x \rightarrow y$ & Host $x$ processing a secure bit exchange with host $y$. \\
    \hline
    $\bigstar$ & Host of this row is being utilized. \\
    \hline
    $\bigcirc$ & Host of this row is inactive. \\    
    \hline
    \end{tabular}
  \end{center}
  \caption{This table is the legend of table~\ref{table:steps}}
  \label{table:legend}
\end{table}

\begin{figure}[H]
    \caption{This is the plot of table~\ref{table:ab}. The data points are plotted along with a linear regression line which is $f(N)=1.3192982456 \cdot N - 1.301754386$, and the coefficient of determination is $R^{2} = 0.988989157$. The horizontal axis is $N$ meaning the number of hosts in the star network. The vertical axis is SBEP($N$) meaning the number of SBEP steps needed for a network with $N$ hosts.}
    \label{fig:linearplot}
  \centering
    \includegraphics[width=1 \textwidth]{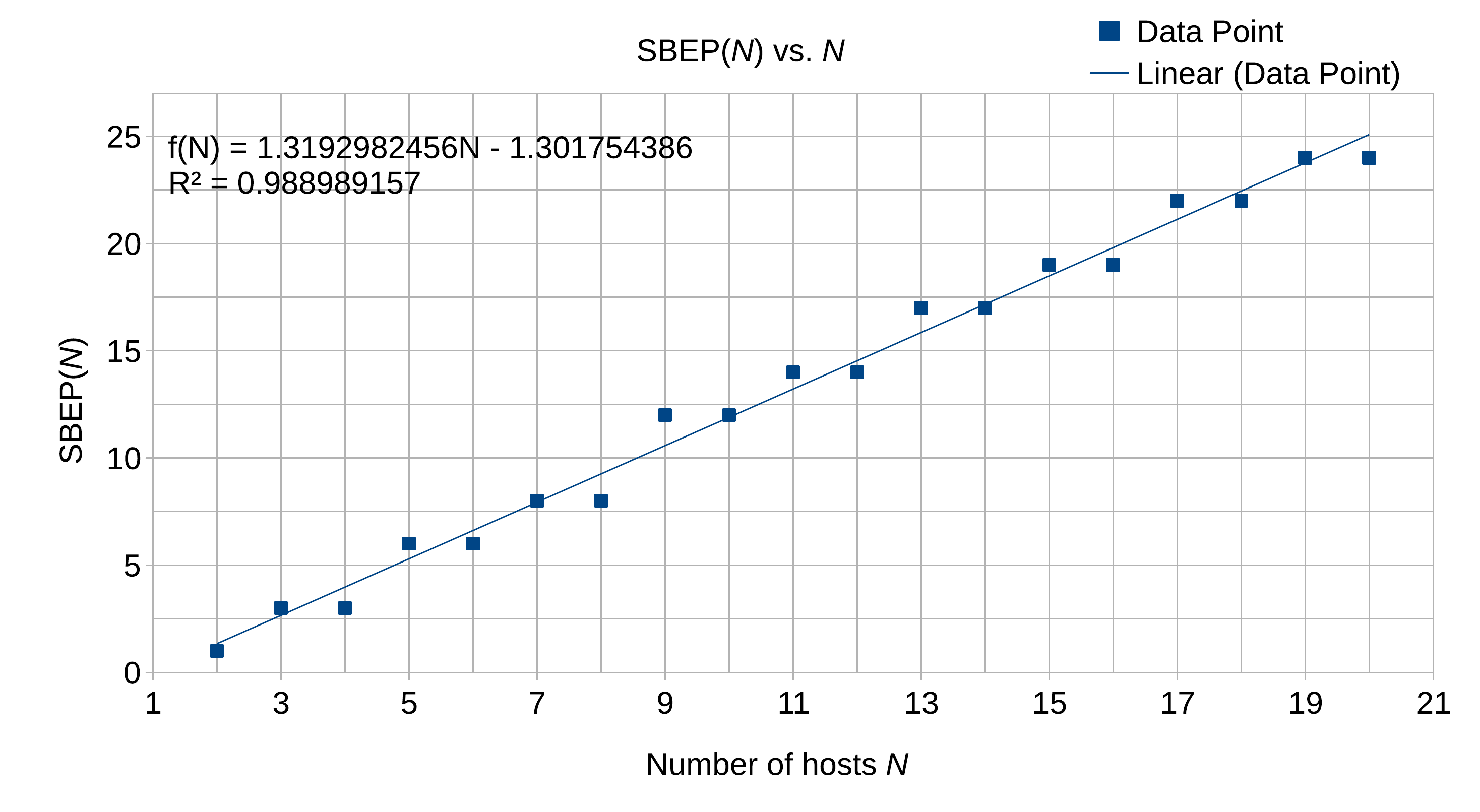}
\end{figure}

The number of SBEPs needed in the $\mathrm{STAR}$ protocol is dependent on the number of hosts $N$ in the network. Table~\ref{table:ab} shows the number of SBEPs needed in the star network for every host to process a secure bit exchange with every other host in the network, for star networks with up to $20$ hosts. Figure~\ref{fig:linearplot} is the plot of table~\ref{table:ab}, with $N$ being the independent variable and SBEP being the dependent variable. The linear regression line is $f(N)=1.3192982456 \cdot N - 1.301754386$, and the coefficient of determination is $R^{2} = 0.988989157$.

\begin{table}[H]
  \begin{center}
    \begin{tabular}{| r | p{6cm} |}
    \hline
    $N$, number of hosts in star network & SBEP($N$), number of SBEP steps needed for a network with $N$ hosts\\
    \hline
    2 & 1 \\
    3 & 3\\
    4 & 3 \\
    5 & 6 \\
    6 & 6 \\
    7 & 8 \\
    8 & 8 \\
    9 & 12 \\
    10 & 12 \\
    11 & 14 \\
    12 & 14 \\
    13 & 17 \\
    14 & 17 \\
    15 & 19 \\
    16 & 19 \\
    17 & 22 \\
    18 & 22 \\
    19 & 24 \\
    20 & 24 \\     
    \hline
    \end{tabular}
  \end{center}
  \caption{This table shows the number of SBEPs needed in star networks with 2 hosts to 20 hosts, with every host in the network to process a secure bit exchange with every other host.}
  \label{table:ab}
\end{table}

The patterns and relations in the star network protocol can be seen in table~\ref{table:ab} and Figure~\ref{fig:linearplot}. Note that when $N$ is evenly divisible by 2 then it will take exactly 2 SBEP steps for every host \textit{i} to process a secure bit exchange with their nearest neighbor host \textit{i}+1. If $N$ is not evenly divisible by 2 then it will take exactly 3 SBEP steps for every host \textit{i} to process a secure bit exchange with their nearest neighbor host \textit{i}+1. The results are the same for every case when $N$ is divided by 3, 4, 5, ..., $(N-1)/2$, and every host \textit{i} processes a secure bit exchange with their second, third, fourth, ..., $(N-2)/2th$ nearest neighbor, that is host \textit{i}+2, \textit{i}+3, \textit{i}+4, ..., \textit{i}+$(N-2)$ respectively. There is a unique case when $N$ is even and is divided by $N/2$, in this case only one SBEP step is needed to process a secure bit exchange. The residual steps are combined whenever possible. For example, in the case when $N=7$, the 6$th$ and 9$th$ steps can be combined into one step resulting in one less SBEP step. These patterns and relations were used to conceive equations~\eqref{subeq1} through~\eqref{subeq4}, where the ``$\lceil{}\rceil$'' symbol in the equations is the ceiling function, $N$ is the number of hosts and $\mathrm{SBEP}(N)$ is the number of SBEPs needed to share an independent secure bit for each possible pairs formed in the network, which means each host share $N-1$ secure bits. (Note, after this sharing, each possible pairs formed in the network has only a single bit of their respective secure key. Thus to share a key with k bits, the above process must be repeated k times.)

\begin{subequations}
 \label{eq1}
 \begin{align}
  \mathrm{SBEP}(N) = N+  \left\lceil{\frac{N}{4}}\right\rceil -2 \mbox{    for } N \leq 8 \mbox{ and } N \mbox{ is even}. \label{subeq1} \\
  \mathrm{SBEP}(N) = N+  \left\lceil{\frac{N}{4}}\right\rceil -1 \mbox{    for } N \leq 8 \mbox{ and } N \mbox{ is odd}. \label{subeq2} \\
  \mathrm{SBEP}(N) = N+  \left\lceil{\frac{N}{4}}\right\rceil -1  \mbox{    for } N > 8 \mbox{ and } N \mbox{ is even}. \label{subeq3} \\
  \mathrm{SBEP}(N) = N+  \left\lceil{\frac{N}{4}}\right\rceil  \mbox{    for } N > 8 \mbox{ and } N \mbox { is odd}. \label{subeq4}
 \end{align}
\end{subequations}

The reliability of the star network is dependent on its center switch, cable, and key exchanger. One could sabotage the entire network just by damaging the center switch in the star network. If a cable or key exchanger is damaged in the star network then the affected host will be effectively disconnected from the entire network, but the unaffected hosts will be able to continue processing a secure bit exchange with other hosts in the network. 

To add additional hosts in the star network will require every hosts in the network to change the protocol from $N$ to $N+1$, which is a relatively simple process since the protocols can be preprogrammed in the hosts.

The star network could be utilized in many situations including vehicle information networks \cite{c181, c185} and inside equipment with components spread around a central processing unit such as a computer.

%
%
%
%
%
%
%

\subsection{Comparing network topologies}

Table~\ref{table:comp} compares the fully connected network with $N-1$ key exchangers per host denoted by $\mathrm{FCN}_{N-1}$, the fully connected network with $1$ key exchanger per host denoted by $\mathrm{FCN}_{1}$, the linear chain network protocol with $2$ key exchangers per hosts is denoted by $\mathrm{LCH}$, and the star network protocol with $1$ communicator per host denoted by  $\mathrm{STAR}$. As can be seen from table~\ref{table:comp} the fastest network is the $\mathrm{FCN}_{N-1}$ network, the networks with the least cost of cables are the linear chain network and the star network, and the networks with the least cost of key exchangers are $\mathrm{FCN}_{1}$, linear chain network, and star network. These results will hold for both KLJN and QKD systems. These results show that the star network has better performance than the linear chain network with similar cost of cables and key exchangers.

\begin{table}[h]
  \begin{center}
    \begin{tabular}{| c | c | c | c |}
    \hline
    Network type & $T_\mathrm{cable}(N)$ & $T_\mathrm{ke}(N)$ & $T_\mathrm{time}(N)$\\
    \hline
    $\mathrm{FCN}_{N-1}$ & $O(N^{2})$ & $O(N^{2})$ & $O(1)$ \\
    $\mathrm{FCN}_{1}$ & $O(N^{2})$ & $O(N)$ & $O(N)$ \\
    $\mathrm{LCH}$ & $O(N)$ & $O(N)$ & $O(N^{2})$ \\
    $\mathrm{STAR}$ & $O(N)$ & $O(N)$ & $O(N)$ \\
    \hline
    \end{tabular}
  \end{center}
  \caption{This table summarizes the complexities of the fully connected networks $\mathrm{FCN}_{N-1}$ and $\mathrm{FCN}_{1}$, the linear chain network protocol $\mathrm{LCH}$, and the star network protocol $\mathrm{STAR}$.}
  \label{table:comp}
\end{table}

The robustness and reliability of each network is dependent on its geometric topology. If a cable is damaged then it is best to have a $\mathrm{FCN}_{N-1}$ network since only one connection between two hosts will be lost. In the linear chain network the entire network will be divided. In the star network the affected host will be completely disconnected from the network. If a key exchanger is damaged then it is best to have a linear chain network since the only consequences will be a slower secure bit exchange process, but every host will still be able to process a secure bit exchange with every other host. In the $\mathrm{FCN}_{N-1}$ network a damaged key exchanger will only affect one connection between two hosts. In the star network a damaged key exchanger will completely disconnect the affected host from the entire network. Another weakness of the star network is the center switch, if the center switch is damaged then the entire network is disconnected. Based on these three networks one can argue that the most robust reliable network is the $\mathrm{FCN}_{N-1}$ followed by the linear chain network, and the least robust network of these three would be the star network.

To add hosts to the $\mathrm{FCN}_{N-1}$ network would be trivial since the $\mathrm{FCN}_{N-1}$ does need a protocol, all that is needed is to connect the host to every other host. To add hosts to the linear chain network and the star network will require every hosts in the network to change the protocol from $N$ hosts to $N+1$ hosts, this will be a relatively simple process as every host can be preprogrammed.

\subsection{Open questions and future studies}

The star network has complexity of $O(N)$ for the number of cables, key exchangers, and time, but there are still numerous other geometric network topologies that have not been explored that might benefit KLJN and QKD systems. Other examples for possible networks include matrix networks, that is a grid of several vertical lines and horizontal lines. The matrix network might be a good model for an urban city with squared blocks. A wheel network is another possibility that might outperform the star network. A wheel network is similar to a star network but with a connecting loop around the branches. A web network is another interesting network similar to the wheel network but with concentric circles connecting the inner branches. A web network is similar to a spider web with each node being a host. A cube network is another interesting possibility that could be utilized in a skyscraper. A cube network is similar to the matrix network except that it is three dimensions. A sphere network might be another interesting three-dimensional network that can be compared with the cube network.

Since different geometrical topologies give different trade-offs, another interest is to explore the trade-offs of the different networks, and why it is preferable to sacrifice speed, communicators, or key exchangers for infrastructure and vice versa. Another possible interest is to analyze and compare every geometric network with different number of communicators and how well they scale with speed. Another possibility is to combine several of these networks into one network and analyze its performance, in graph theory this is known as hybrid networks.

Different geometic network structures have different vulnerabilites, an analysis of each network's vulnerabilites, robustness, reliability, and different kinds of attacks would be interesting to explore and compare.

\section{Conclusions}

In this study we considered the need for unconditional secure key exchange along with the need to have P2P networks since QKD and KLJN require P2P networks. We reviewed a simple P2P network known as the fully connected network. We also reviewed the linear chain network and analyzed the star network to compared it with fully connected networks and the linear chain network. We conceived a protocol and equations~\eqref{subeq1} through~\eqref{subeq4} to describe the star network. The results show that the star network compares favorably to the linear chain network and the fully connected network. Even though the star network utilizes only one key exchanger per host, its time complexity is superior to that of the linear chain network, while its cable complexity is the same. The star network's cable and key exchanger complexity is superior to that of the fully connected network, while its time complexity is worse than $\mathrm{FCN}_{N-1}$, but is similar to $\mathrm{FCN}_{1}$. We found that the star network fairs worse than the linear chain network and the fully connected network in robustness and reliability as the star network can be entirely disconnected by damaging the center switch. We then considered several other possible network geometries that might be interesting to explore and to compare.




%


%

\bibliographystyle{mdpi}
\makeatletter
\renewcommand\@biblabel[1]{#1. }
\makeatother



%


%

\end{document}